%% file: main.tex
\documentclass[sigconf]{acmart}
\usepackage{booktabs} 

\usepackage{graphicx}
\usepackage{balance}  

\usepackage[linesnumbered,algoruled,boxed,lined]{algorithm2e}
\newcommand{\VALMOD}{{\textit{VALMOD} }}
\newcommand{\VALMAP}{{\textit{VALMAP} }}

\usepackage{comment}
\usepackage{url}

\usepackage{amsmath}

\renewcommand\footnotetextcopyrightpermission[1]{} 
\begin{document}

\title[Suite for Easy Detection of Motif in Data Series]{VALMOD: A Suite for Easy and Exact Detection of\\ Variable Length Motifs in Data Series}


\author{ Michele Linardi}
\affiliation{%
	\institution{ LIPADE, Paris Descartes University}
}
\email{michele.linardi@parisdescartes.fr}
\author{Yan Zhu}
\affiliation{%
	\institution{UC Riverside}
}
\email{yzhu015@ucr.edu}
\author{Themis Palpanas}
\affiliation{%
	\institution{LIPADE, Paris Descartes University}
}
\email{themis@mi.parisdescartes.fr}
\author{Eamonn Keogh}
\affiliation{%
	\institution{UC Riverside}
}
\email{eamonn@cs.ucr.edu}

\begin{abstract}
Data series motif discovery represents one of the most useful primitives for data series mining, with applications to many domains, such as robotics, entomology, seismology, medicine, and climatology, and others. The state-of-the-art motif discovery tools still require the user to provide the motif length.  
Yet, in several cases, the choice of motif length is critical for their detection. 
Unfortunately, the obvious brute-force solution, which tests all lengths within a given range, is computationally untenable, and does not provide any support for ranking motifs at different resolutions (i.e., lengths).     
We demonstrate VALMOD, our scalable motif discovery algorithm that efficiently finds all motifs in a given range of lengths, and outputs a length-invariant ranking of motifs. 
Furthermore, we support the analysis process by means of a newly proposed meta-data structure that helps the user to select the most promising pattern length. This demo aims at illustrating in detail the steps of the proposed approach, showcasing how our algorithm and corresponding graphical insights enable users to efficiently identify the correct motifs. (Paper published in ACM Sigmod Conference 2018.)
\end{abstract}

%
%




\maketitle

\input{intro}
\input{sec2}

\input{sec3}

\input{sec4}
\input{conclusions}


\bibliographystyle{ACM-Reference-Format}
\bibliography{GeneralBIB}  
\end{document}

%% file: intro.tex
\section{Introduction}

\noindent{\bf State of the art motif discovery.} 
Over the last decade, data series\footnote{If the dimension that imposes the ordering of the series is time, then we talk about \emph{time series}. However, a series can also be defined through other measures (e.g., angle in radial profiles in astronomy, mass in mass spectroscopy, position in genome sequences, etc.). Throughout this paper, we will use the terms \emph{time series}, \emph{data series}, and \emph{sequence} interchangeably.} motif discovery has emerged as perhaps the most used primitive for data series data mining, and it has many applications to a wide variety of domains~\cite{Whitney,DBLP:conf/kdd/YankovKMCZ07}, including classification, clustering, and rule discovery. More recently, there has been substantial progress on the scalability of motif discovery, and now massive datasets can be routinely searched on conventional hardware~\cite{Whitney}. 
The state-of-the art algorithm \cite{DBLP:conf/icdm/ZhuZSYFMBK16} only requires the user to set a single parameter, which is the desired length of the motifs. Moreover, the motif mining is supported by the \textit{Matrix profile} output, which is a meta data series storing the z-normalized Euclidean distance between each subsequence and its nearest neighbor. The Matrix profile does not exclusively provide the motif, i.e., the subsequence pair with the smallest distance, but also permits to rank and filter out the other pairs, giving also a convenient and graphical representation of their occurrences and proximity. In order to categorize motifs, we call the $k$ subsequences, with the $k$ smallest best match distances, top-$k$ motif pairs.      

\noindent{\bf Motif discovery of different lengths.} Exact Motif discovery has merely become a single input parameter problem, namely the length of the patterns we want to mine.
Unfortunately, this  technique comes with an important lack. It does not provide an effective solution for trying several motif length in a range.
If one has no cues about an effective fixed length, the simplest solution would be to run the algorithm over all lengths in the range and rank the various motifs discovered, picking eventually the patterns, which contain the desired insight. 
Clearly, this possibility is not optimal for at least two reasons; the scalability, since finding motif of one fixed length takes $O(n^2)$ time, and also because it does not provide an effective way to compare motifs of different lengths.
In this work, we demonstrate the solution to this problem, we recently introduced in \cite{VALMOD}, to mine Motif discovery of variable lengths.
In our contribution we propose \VALMOD, the first approach for mining top-$k$ motif pairs of variable length, which is up to orders of magnitude faster/more scalable than the alternatives that have been proposed in the literature.

\begin{figure*}[tb]
	\centering
 	\includegraphics[trim={0cm 11cm 2cm 3.8cm},scale=0.65]{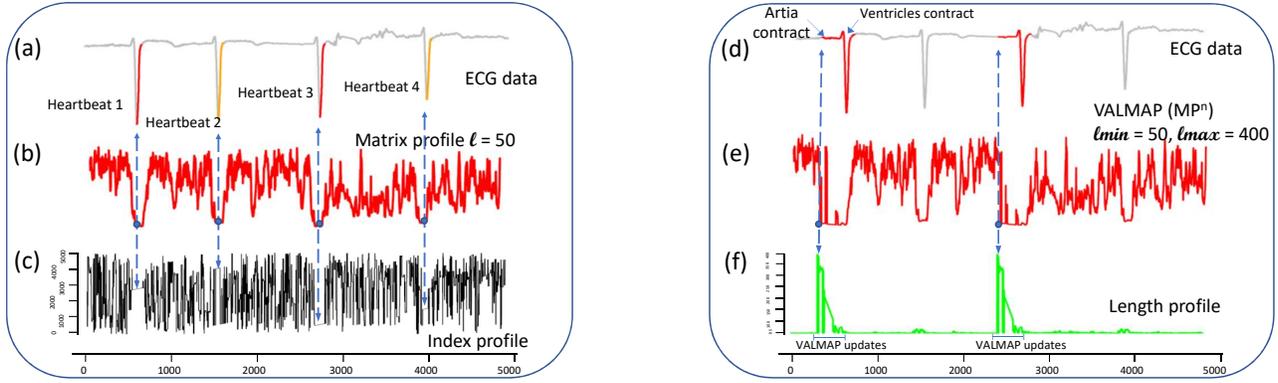}
 	\caption{\textit{Left}) \textit{(a)} Snippet of ECG recording with highlighted motifs of length \textit{50}, \textit{(b)} Matrix profile computed with subsequence length \textit{50}. \textit{(c)} Index profile, reporting the offsets of the best match. \textit{Right}) \textit{(d)} Snippet of ECG recording with highlighted motifs of length \textit{400}, \textit{(e)} VALMAP $MP^n$, \textit{(f)} VALMAP Length profile.}
 	\label{mpECG}
 	\vspace*{-0.3cm}	
\end{figure*}

In order to show the superiority of variable-length motif discovery, consider the following example.
In Figure~\ref{mpECG} (left) swe depict a snippet of an Electrocardiogram (ECG) recording in (a), paired with its Matrix profile, computed with fixed subsequence length: $\ell=50$ in (b). 
Note that each value in the Matrix profile corresponds to a point in the data, which is the representative starting point of a subsequence of length $\ell$.
Hence, given a data series $D$ of length $|D|$, a Matrix profile records $|D|-\ell+1$ distances, avoiding trivial matches \cite{VALMOD}. 
In Figure~\ref{mpECG}.(c) we plot the Index profile, which contains the offsets of the best matches.

Looking at the Matrix profile in this example, we note four deep valleys, which suggest the presence of very close matches, namely the motifs. Starting from the Matrix profile, it suffices to follow the dotted lines upwards, in order to detect the motifs, and downwards for finding the position of each subsequence best match.   
Despite the motifs (heartbeats) are easily detectable \textit{to the naked eye}, since the snippet is relatively short, the highlighted motifs in Figure~\ref{mpECG}.(a) (red/orange subsequences), just report the second half of a ventricular contraction, giving thus a partial and unsatisfactory result. 


In the next section we present the complete details of the \VALMOD algorithm.

%% file: sec2.tex
\section{VALMOD Motif Management}
\label{sec:VALMOD Motif Management}
\begin{figure}[tb]
	\centering
	\includegraphics[trim={1cm 0cm 0cm 4.8cm}, scale=0.32]{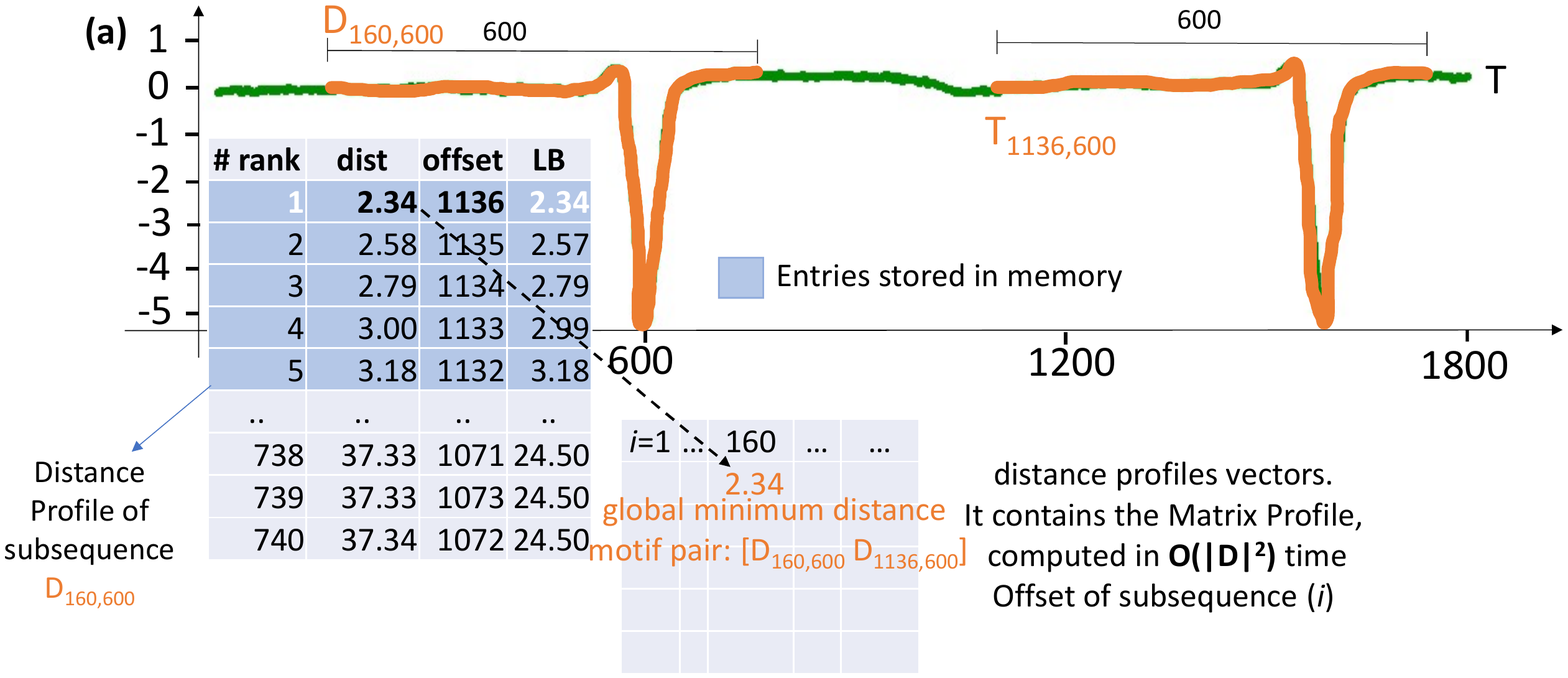}
	\includegraphics[trim={2cm 7cm 0cm 9.3cm}, scale=0.32]{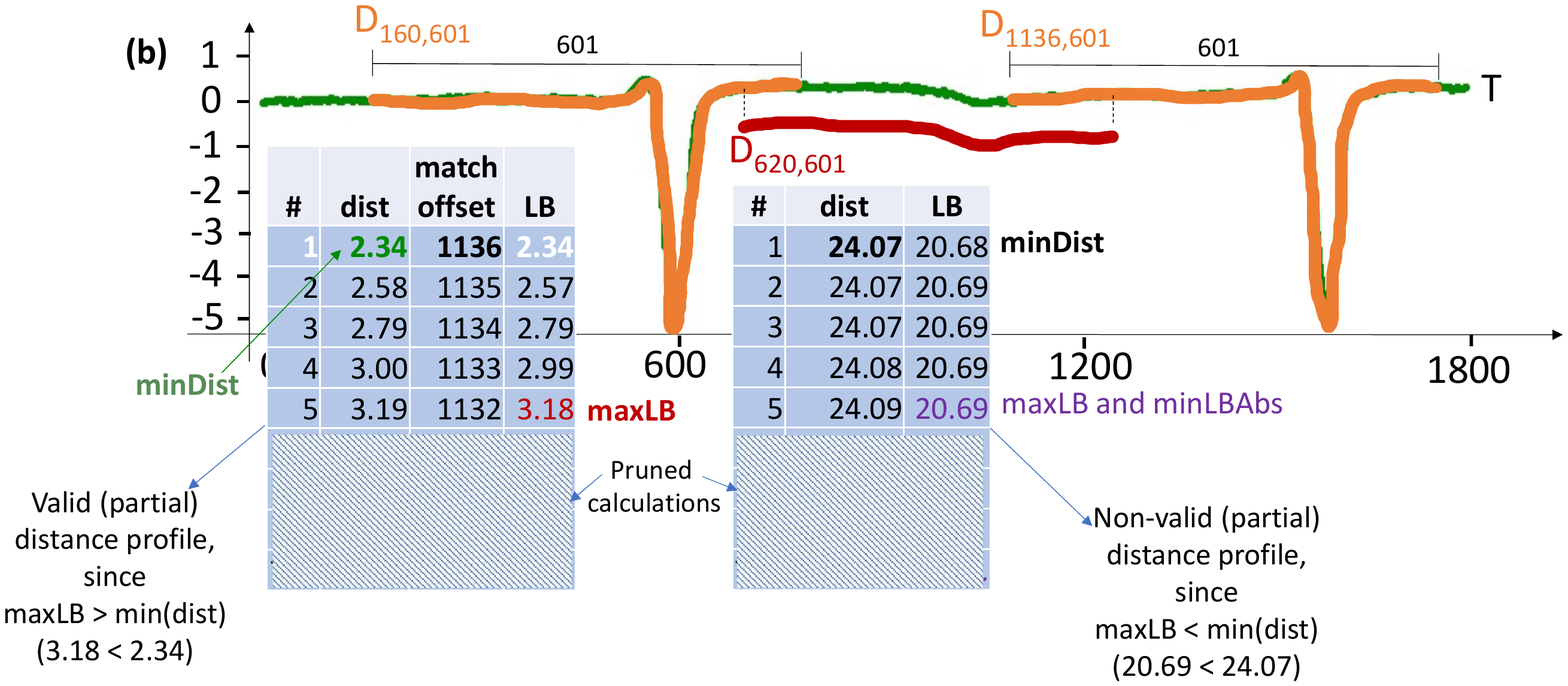}
	\caption{ \textit{(a)} ECG snippet with distance profile of subsequence $D_{160,600}$, \textit{(b)} Partial distance profiles computation for length \textit{601}.}
	\label{ExampleVALMOD} 
	\vspace*{-0.4cm}
\end{figure}

\noindent{\bf VALMOD algorithm} As previously introduced, our algorithm, VALMOD (Variable Length Motif Discovery), given a data series $D$, starts by computing the Matrix profile using the smallest subsequence length, namely $\ell_{min}$, within a specified input range $[\ell_{min},\ell_{max}]$.
The key idea of our approach is to minimize the work that needs to be done for succeeding subsequence lengths ($\ell_{min}+1$, $\ell_{min}+2$, $\ldots$, $\ell_{max}$).
To explain the main components and the idea of our algorithm we present a short example in Figure~\ref{ExampleVALMOD}.

We start to consider the data series $D$ in (a) (snippet of ECG recording). To compute the Matrix profile, \VALMOD considers all the contiguous subsequences of length $\ell{min}$, computing for each one the \textit{Distance profile} in $O(|D|)$ time. This latter, contains the z-normalized Euclidean distance between a subsequence and all the other in $D$. In Figure~\ref{ExampleVALMOD}.(a) we report a distance profile for the subsequence $D_{160,600}$ (the subscript denotes offset=\textit{160} and length=\textit{600}). The minimum distance of each distance profile is a point of the Matrix profile.

We moreover introduce a new lower bounding distance\cite{VALMOD}, which lower bounds the true Euclidean distances between longer subsequences in the distance profiles. 
We initially compute this lower bound from scratch, using as a base the true Euclidean distances computation of subsequences with length \textit{600}. For the larger lengths, we update the lower bound, considering only the variation generated by the trailing points in the longer subsequences. This measure enjoys an important property: if we rank the subsequences according to this measure (ascending order), the same rank will be preserved along all the lower bound updates. 
We want to exploit this property, in order to prune computation. Hence, when the distance profiles are computed (in this example for length=\textit{600}), we keep in memory the $p$ Euclidean distances, which have the smallest lower bounding distance (LB); this is done for each distance profile. 
We show in Figure~\ref{ExampleVALMOD}.(b)  how the algorithm proceeds for the length \textit{601}. Instead of computing from scratch the whole distance profiles, we consider just the elements we stored in the previous step. Here, each distance profile is denoted as \textit{partial distance profile}. We proceed computing the true Euclidean distances of each partial distance profile, updating the relative LB (this result is depicted in Figure~\ref{ExampleVALMOD}.(b). After this operation, we may have two cases: if in a new computed distance profile the minimum true distance (\textit{minDist}) is shorter than the maximum lower bound (\textit{maxLB}), we know that no elements, among those not computed, can be smaller than minDist. In this case a partial distance profile becomes a \textit{valid distance profile}, as in the case of the subsequence $D_{160,601}$.
On the other hand, when \textit{maxLB} is smaller than \textit{minDist}, as in the case of subsequence $D_{620,601}$, no true minimum distance is found within the distance profile.
At the end of this process, we pick the minimum \textit{maxLB} of all the non-valid distance profile, which is denoted as \textit{minLBAbs}. Hence, all the \textit{mindist} in the valid (parital) distance profiles, smaller than \textit{minAbsLB} are considered top-$k$ motif distances.
If no \textit{mindist} are smaller than \textit{minAbsLB}, we recompute only the distance profiles, which have the \textit{maxLB} smaller than the smallest \textit{mindist} found, since only those may contain better matches than the already computed ones.  
We keep extracting in this way, the top-$k$ motifs of each length, until $\ell{max}$.

\begin{figure}[tb]
	\includegraphics[trim={2cm 6.5cm 1cm 0cm},scale=0.36]{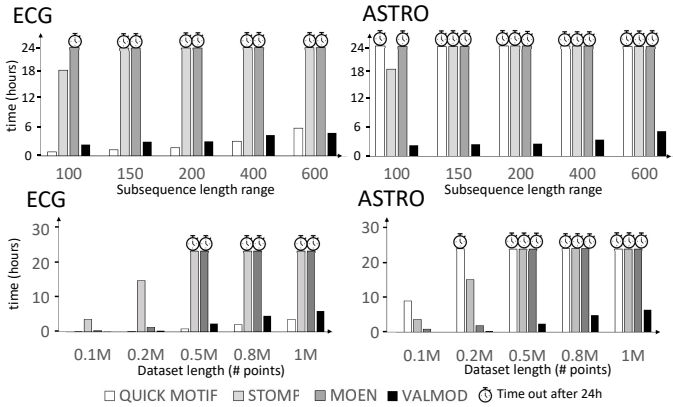}
	\caption{ \textit{(top)} Time over motif length ranges (default $\ell_{min}$=\textit{1024}, data series length= \textit{0.5M}. \textit{(bottom)} Time over series length (default length range=\textit{100}). }  
	\label{Scalability1}
		\vspace*{-0.5cm}
\end{figure}

\noindent{\bf Experimental Evaluation.} To benchmark \VALMOD, we use several different datasets in \cite{VALMOD}, comparing it with two types of algorithms.
The first are two state-of-the-art motif discovery algorithms, which receive a single subsequence length as input: QUICKMOTIF~\cite{DBLP:conf/icde/LiUYG15} and STOMP~\cite{DBLP:conf/icdm/YehZUBDDSMK16}. 
In our experiments, they have been adapted to find all the motifs for a given subsequence length range. 
The other approach in the comparative analysis is MOEN~\cite{DBLP:journals/kais/MueenC15}, which accepts a range of lengths as input, producing the best motif pair for each length. 
We report in Figure~\ref{Scalability1} a sample of the experiments we conducted (detailed experimental results on several datasets are reported elsewhere~\cite{VALMOD}). 
Here, we show the results of \VALMOD, which finds motifs in an Electrocardiogram recording (ECG) and in a data series representing celestial objects (ASTRO)~\cite{VALMOD}. We couple the \VALMOD results with those of its competitors. In the plots, we report the total execution time of \VALMOD, which includes all the operations performed by the algorithm (also the \VALMAP computation introduced later), varying motif length ranges (Figure~\ref{Scalability1} (top)) and the size of the input data series, considering different prefix snippets (Figure~\ref{Scalability1} (bottom)). 
From this experiment, we observe that VALMOD maintains a good and stable performance across datasets and parameter settings, quickly producing results, even in cases where the competitors do not terminate within a reasonable amount of time.

\noindent{\bf Rank Motif Pairs of Variable Lengths.} Since we can discover motifs of different lengths, we propose a ranking method, suitable for comparing different-length patterns.
We aim to favor longer and similar sequences in the ranking process of matches that have different lengths. 
As a consequence, we factorize the Euclidean distance by the following quantity: $sqrt(1/\ell)$, where $\ell$ is the length of the sequences. We call the new distance, \textit{length normalized distance}~\cite{VALMOD}.

\noindent{\bf VALMAP.} While the proposed motif rank weights the subsequences importance according to the ratio distance-length, we want to know also, whether and how the motif pairs changes, helping the user to extract the desired insights at the correct length. 
To that extent, we introduce a new meta-data, called Variable Length Matrix Profile (\textit{VALMAP}), maintaining the same logic and structure of the Matrix profile depicted in Figure~\ref{mpECG} (top), with the difference that this new structure carries length normalized distances and it is coupled with a new vector called \textit{Length profile}, which contains the lengths of the subsequences.
More formally, given a data series $D$, and a range of subsequence lengths, whose extremes are denoted by $\ell_{min}$ and $\ell_{max}$, we define \textit{VALMAP} as a triple $\langle MP^{n} \in \mathbb{R}^{|D|-\ell_{min} + 1}, IP \in \mathbb{N}^{|D|-\ell_{min} + 1},  LP \in \mathbb{N}^{|D|-\ell_{min} + 1} \rangle$, where $MP^{n}$ is the Matrix profile containing length normalized distances, whereas $IP$ and $LP$ are the relative Index and Length Profile.
If we consider just a fixed length, \VALMAP will coincide with the length normalized version of the Matrix profile, with a flat Length profile.
This is basically the structure that \VALMOD builds, considering subsequences of length $\ell_{min}$.
In the second stage, we can update \VALMAP using the top-$k$ motif pairs, computed for each length until $\ell_{max}$. 
We thus consider each ($D_{i,\ell_{min}+1} , D_{j,\ell_{min}+1} $) $\in$ top-$k$ motif pairs, where $i,j$ are the subsequences offsets, $\ell_{min}+1$ their lengths and $d^{n}_{i,j}$ their length normalized Euclidean distance. Note that in a motif pair the right subsequence is the one with the absolute shortest distance to the one at the left. 
Hence, \VALMAP, $MP^{n}[i]$ is updated with $d^{n}_{i,j}$ if $d^{n}_{i,j} < MP^{n}[i]$, which was containing the distance between $D_{i,\ell_{min}}$ and its best match. If this update takes place, the Index and Length profile are respectively assigned with $j$, the offset of the new best match, and $\ell_{min}+1$ the new length.
The update operation takes place for each top-$k$ motif pair of any length between $\ell_{min}$ and $\ell_{max}$.
Once the algorithms ends, \VALMAP contains a picture of the motif pairs showing, at which length the last update takes place. If a motif pair is updated, this implies that a longer pattern represent a better match and thus it might reveal either a new event or the same event lasting longer.

\noindent{\bf Example of VALMAP Expressiveness.}

In order to show the expressiveness of \VALMAP, we ran \VALMOD on the ECG data snippet previously considered, showing the \VALMAP structure in Figure~\ref{mpECG} (right). We use the following input parameter: $\ell_{min}=50$ and $\ell_{max}=400$.  
We note that \VALMAP reports the motif with the shortest length normalized distance of length \text{56}, which is the same partial event detected by the Matrix profile in the fixed length case, at the top of the picture. 

If we look at the Length profile in Figure~\ref{mpECG}.(f), we observe that, at an earlier time than the discovered motifs pair, a sequence of contiguous updates took place, as we reported. The subsequences concerned have distances almost as short as the one of the best motifs in \VALMAP~, thus, remaining longer and possibly valid matches.

In Figure~\ref{mpECG}.(d) we depict and highlight the motif pair of length \textit{400}. Immediately, we can note that, the subsequences in red, which compose this motif, are a better representation of a recurrent heartbeat. In fact, the two typical components (\textit{Artia and Ventricles contract}) are correctly detected.

\begin{figure}[tb]
	\centering
	\includegraphics[trim={5cm 5.5cm 16cm 4cm},scale=0.45]{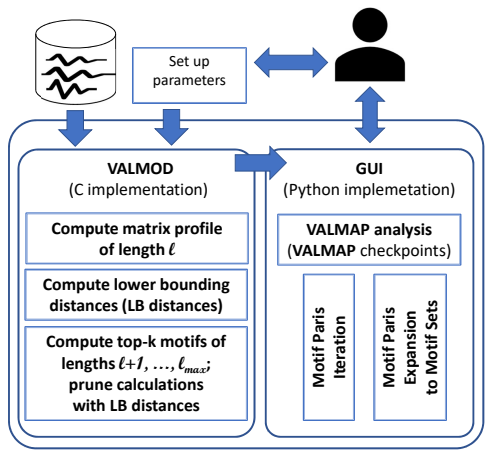}
	\caption{Architecture of VALMOD system. }
	\label{fig:archFlow}
		\vspace*{-0.3cm}
\end{figure}


%% file: sec3.tex
\section{System Description}

\begin{figure}[tb]
	\centering
	\includegraphics[trim={0cm 10cm 21cm 3cm},scale=1.10]{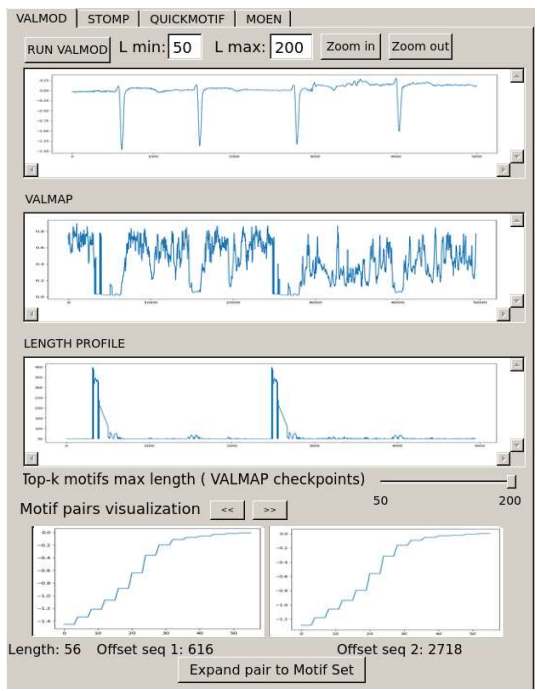}
	\caption{GUI interface showing the interaction with VALMAP. }
	\label{fig:screenShot}
	\vspace*{-0.3cm}
\end{figure}

We now describe the architecture of our system, depicted also in Figure~\ref{fig:archFlow}.
The input is represented by a data series of interest.
As a starting point, the user has the possibility to inspect the data and also setting the desired parameter (lengths range [$\ell_{min}$,$\ell_{max}$]).
Afterwards, she can run the \VALMOD algorithm, which is a part of the system back-end we implemented in C. 
Once terminated, \VALMOD outputs the \VALMAP meta-data. This latter is thus sent to the front-end, implemented in Python. 
Here, the user can interact with the system analyzing the showcased elements, such as:
\begin{itemize}
	\item the checkpoints of the VALMAP, namely all the updates occurred from the length $\ell{min}$ till the desired length, selected with a dedicated slider. 
	\item all the top-$k$ motifs of variable length, which \VALMAP reports. 
	\item expand a selected motif pair to the relative Motif Set, containing all the similar subsequences of the pair in the data. 
\end{itemize}
In Figure~\ref{fig:screenShot} we show a screen-shot of the VALMAP analysis in our demonstration.

%% file: sec4.tex
\section{Demonstration}

We now present the scenarios proposed to the audience.
\noindent{\bf Need for Variable Length Motifs.} 
We will showcase variable length motif discovery using \VALMOD on different real datasets~\cite{VALMOD}, including ECG and ASTRO, as well as 
datasets coming from the domains of \textit{Entomology} and \textit{Seismology}. 
In these two particular cases, the user can understand the importance of using variable length motif detection (with the support of \VALMAP), in order to identify patterns of interesting behavior exhibiting themselves as sequences of different lengths.

\noindent{\bf Traditional Motif discovery VS VALMOD.} In this scenario, we will challenge the user to find the motifs without having any knowledge of their lengths, just by inspecting the data themselves When this takes place, the user can experience the VALMOD support in finding motif pairs that can be of variable length, understanding the quantity and quality of the insights that are not achievable with a simple raw data visual analysis.

\noindent{\bf VALMOD VS Competitors.} In this scenario, the user can compare \VALMOD to alternative approaches used for motif discovery. 
Specifically the audience will note the performance improvement, concerning fixed and variable length motif discovery, and the increased expressiveness provided by \VALMAP.

%% file: conclusions.tex
\section{Conclusions}
\label{sec:conclusions}
In  this  work,  we  present \VALMOD,  a  system  that  can  efficiently  find data series motif of variable length.
As opposed to the other approaches, our framework provides a new meta data-series\\(\VALMAP), which ranks motif pairs of variable length, using a new length normalized distance. 
Our system provides enriched insights, which help to detect not only the correct resolution (length) of an interesting event, but also the occurrences of repeated patterns with different meanings, which are typical in numerous domains.